\documentclass[11pt]{article}
\usepackage{times}
\usepackage{geometry}
\usepackage[utf8]{inputenc}
\usepackage{enumitem,amssymb}
\usepackage{graphicx}
\usepackage{fancyhdr}
\usepackage{aas_macros}
\usepackage{mdframed} 

\usepackage[authoryear]{natbib}
\setcitestyle{authoryear,open={(},close={)}}

\mdfdefinestyle{theoremstyle}{
innertopmargin=\topskip,}
\mdtheorem[style=theoremstyle]{lrptextbox}{}

\newcommand{\tcm}{$21\,\textrm{cm}\,\,$}  

\begin{document}

\title{High-redshift 21cm Cosmology in Canada}
\author{A. Liu, H. C. Chiang, A. Crites, J. Sievers, R. Hlo\v{z}ek}
\date{Submitted to LRP2020 panel on 30th September 2019}

\maketitle

%

\section{Introduction}

Despite recent advances in observational cosmology, much of our high-redshift Universe remains unexplored. Cosmic Microwave Background (CMB) experiments have provided exquisite measurements of the initial conditions of our Universe, while galaxy surveys and traditional astronomical measurements have provided a reasonably detailed view of the $z < 6$ Universe. Missing from this is everything in between---from the post-recombination era known as the Dark Ages (prior to the formation of the first luminous structures) to Cosmic Dawn (when the first stars and galaxies were formed) to the Epoch of Reionization (EoR; when the first galaxies systematically ionized the intergalactic medium). In the past decade, a series of first-generation experiments in \tcm cosmology have made considerable progress towards changing this status quo. In the next decade, \tcm cosmology has the potential to become a workhorse probe of the rich astrophysical and cosmological phenomena residing in the $z>6$ universe.

The key idea behind \tcm cosmology is the abundance of hydrogen in our Universe across a wide range of redshifts, coupled with the existence of the optically thin \tcm hyperfine transition spectral line. This in principle enables radio telescopes to perform a systematic, large-scale mapping of our Universe through the brightness temperature of the \tcm line. Encoded in this line is the density, velocity, ionization state, and spin temperature of hydrogen across a broad range of length scales. This provides a sensitive probe of not only the astrophysics of early galaxy formation, but also of fundamental cosmology.\footnote{As is conventional in the field, in this white paper we use the phrase ``\tcm cosmology" even when we are referring to galaxy formation astrophysics that one might not traditionally deem ``cosmology".} \tcm cosmology builds on the foundations of observational cosmology provided by galaxy surveys and the Cosmic Microwave Background (CMB) along several frontiers 
\citep{tegmarkzaldarriaga2009}:

\begin{itemize}
    
    \item {\bf The redshift frontier.} The \tcm line is one of the only probes of our Universe at redshifts between recombination and the formation of the first luminous objects. Opening new redshift windows will provide direct constraints on the properties of these objects, and may also provide tantalizing hints of new physical phenomenology (see Section \ref{sec:landscape}).
    
    \item {\bf The sensitivity frontier.}  Hydrogen is omnipresent in our Universe, allowing large volumes to be mapped in 3D, with the line-of-sight distance automatically given by the redshift of the line. The \tcm line can in principle access orders of magnitude more cosmological modes than galaxy surveys or the CMB \citep{Stage2}.
    
    \item {\bf The scale frontier.} 
    The \tcm line can also access small scales. Unlike the CMB, small scales are not Silk damped, allowing measurements down to the Jeans scale. Additionally, at high redshifts the modes remain linear to smaller scales, simplifying theoretical analyses. Finally, the wide-field nature of low-frequency observations allow access to the largest scales.
    
\end{itemize}

\section{The Current Landscape for $21\,\textrm{cm}$ Cosmology at $z>6$}
\label{sec:landscape}

Of course, any observational probe is only as successful as our ability to control experimental systematics and achieve design sensitivities. In \tcm cosmology, this is a challenging problem, although current-generation experiments have made considerable progress. Chief amongst the challenges is the problem of foreground contamination. At redshifts $z>6$, the \tcm line falls into the low-frequency radio regime, where (for example) Galactic synchrotron radiation is orders of magnitude brighter than the cosmological signal that one seeks. A large dynamic range is therefore required for any measurement of the cosmological signal, which in turn places extremely stringent requirements on systematic effects such as mismodellings of one's telescope beams or cable reflections in one's hardware.

Because \tcm cosmology is so technically challenging, there has yet to be a \emph{definitive} detection of the cosmological signal at $z > 6$. However, in the case of the \tcm \emph{global} signal (where one averages the \tcm emission over all angles of the sky to produce a single curve as a function of redshift), there has been a claimed tentative detection by the Experiment to Detect the Global EoR Signature (EDGES). Like most other global signal experiments, EDGES is a well-calibrated single-antenna experiment placed in a radio-quiet environment. In \citet{EDGES2018}, the EDGES team reported a \tcm absorption trough at $z \sim 17$. This was a remarkable claim because the detailed shape of the trough was unexpected, both in its timing and depth (see Section \ref{sec:Frontiers} for more discussion).


Beyond global signal experiments like EDGES are efforts targeting the spatial fluctuations of the \tcm line. To this end, large interferometric facilities with unparalleled digital processing capacities, enormous collecting areas, and unprecedented sensitivities have been built and are now in operation. These include the Giant Metrewave Radio Telescope Epoch of Reionzation project (GMRT; \citealt{Paciga2013}); Murchison Widefield Array (MWA; \citealt{Bowman2013, Tingay2013}), the Low Frequency Array (LOFAR; \citealt{LOFAR2013}), the Donald C. Backer Precision Array for Probing the Epoch of Reionization (PAPER; \citealt{Parsons2010}), the Hydrogen Epoch of Reionization Array (HERA; \citealt{DeBoer2017}), the Owens Valley Radio Observatory Long Wavelength Array (OVRO-LWA; \citealt{Eastwood2018}), and the Large-aperture Experiment to Detect the Dark Age (LEDA; \citealt{Greenhill2012}). While a positive detection of the spatially fluctuating \tcm signal at $z > 6$ remains elusive, considerable progress has been made in the form of increasingly stringent upper limits.

\begin{figure*}[h!]
\centering
\includegraphics[width=1.0\textwidth,trim={0cm 0cm 0cm 0cm},clip]{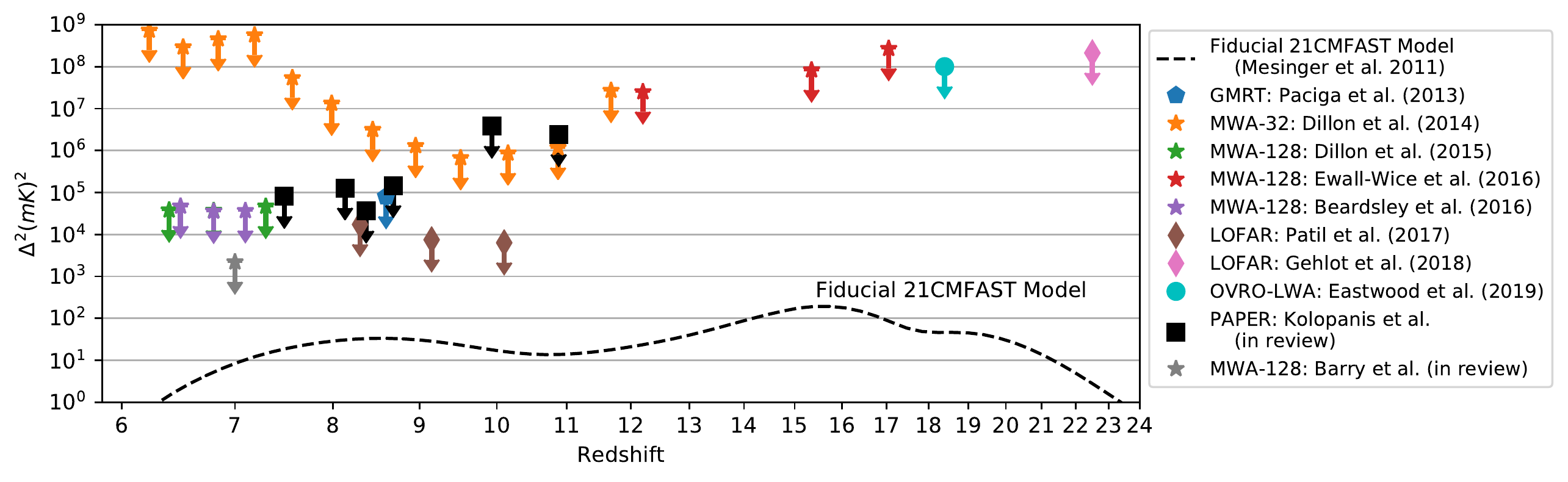}
\caption{A summary of current upper limits from spatial fluctuation experiments, expressed in terms of the ``dimensionless" power spectrum $\Delta^2 (k)$. For each experiment, we pick the $k$ scale for which the results are the most competitive. Except for the LOFAR constraints, this typically means $0.1\,h\textrm{Mpc}^{-1}<k < 1.0\,h\textrm{Mpc}^{-1}$.}
\label{fig:modes}
\end{figure*}

\section{Frontiers in $21\,\textrm{cm}$ Cosmology at $z>6$}
\label{sec:Frontiers}
The next decade will almost certainly see a surge in activity in \tcm cosmology at $z > 6$, driven initially by observations made by instruments that are already built or are nearing completion.

\subsection{Learning from the data} The field of \tcm cosmology will soon enter a new phase, where a large number of funded experiments will have been constructed and will have the sensitivity to make high-significance measurements of spatial fluctuations---in principle. In practice, instrumental systematics and foreground contaminants need to be overcome in order to make this a reality. Fortunately, with the large amounts of data now available, these effects can be studied---and hopefully eliminated---in real data. Already there have been considerable advances in this area (e.g., see \citealt{Kern2019a,Kern2019b} for examples of new techniques for removing the effect of cable reflections and cross-talk in \tcm interferometric data), and this will only improve as datasets get deeper. 
\subsection{A first detection and characterization of spatial fluctuations} 
\begin{figure*}[h!]
\centering
\includegraphics[width=1.0\textwidth,trim={0cm 0cm 0cm 0cm},clip]{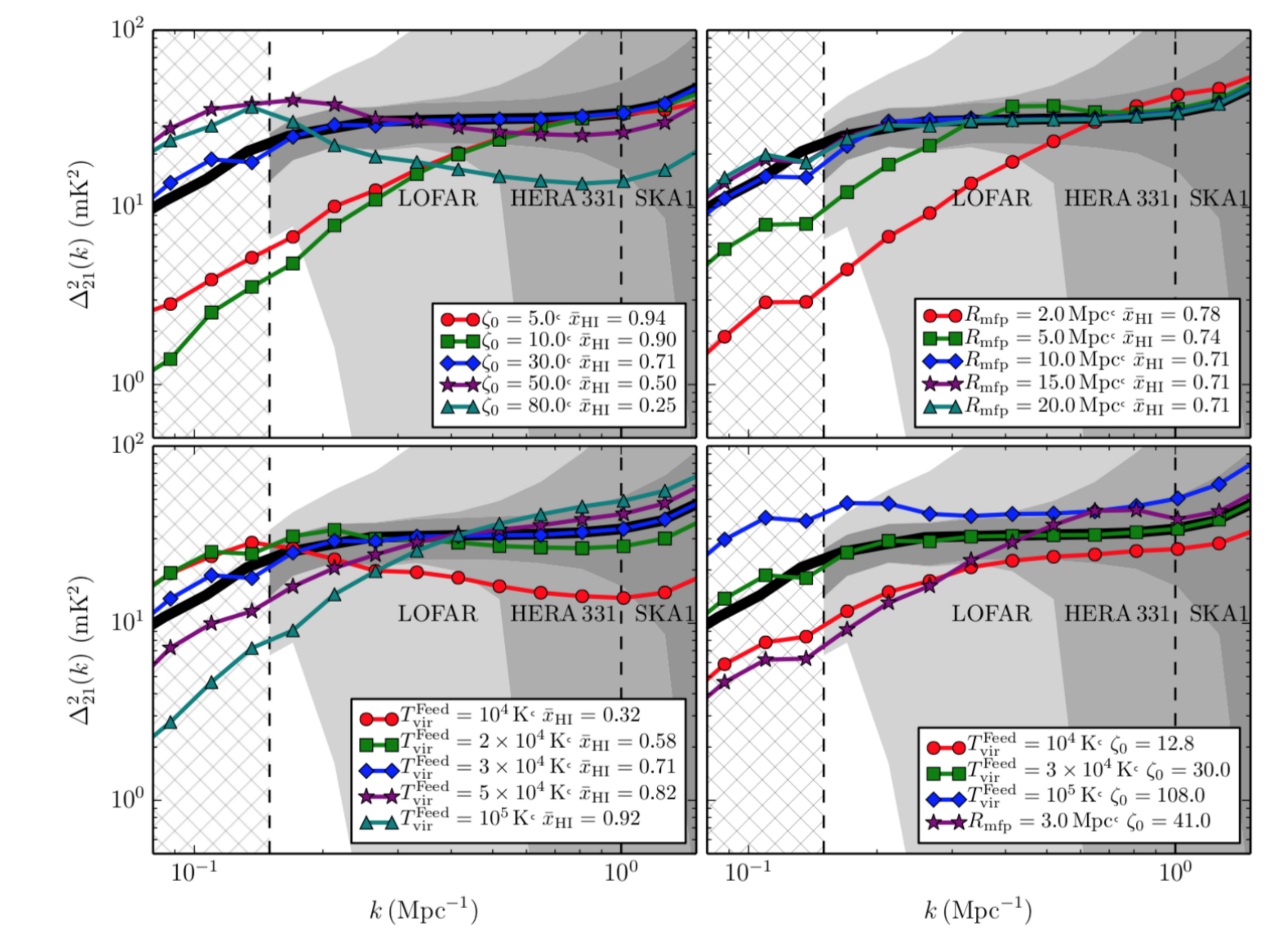}
\caption{Example power spectra at $z =9$, with forecasted sensitivities for LOFAR, HERA, and the SKA. Each panel shows the effects of varying a crucial property of our high-redshift Universe. In the top left, the ionizing efficiency (a measure of the ionizing flux of first-generation galaxies) is varied; in the bottom left, the minimum virial temperature of galaxies; in the top right, the mean free path of UV photons (a measure of IGM absorption); in the bottom right, a combined variation of the aforementioned parameters. From \citet{GreigMesinger2015}.}
\label{fig:pspecs}
\end{figure*}

Should instrumental systematics be surmountable, current instruments like the MWA and HERA possess sufficient sensitivity to make a first positive detection of spatial \tcm fluctuations at $z>6$ within the next few years. With full seasons of observations, these instruments will not simply make a detection, but will perform high-significance \emph{characterizations} of the signal. Figure \ref{fig:pspecs} shows that such measurements will be highly sensitive to the exact nature of first-generation galaxies. This includes parameters that govern the mass function of these galaxies, the nature of their X-ray and UV emission, and the clumpiness of the IGM.

\subsection{Cross-correlations} That the \tcm line is one of the few ways to probe the high-redshift universe is both a blessing and a curse. It is a blessing because it is almost unique in its ability to directly observe the EoR and the immediately preceding epochs. It is a curse in that there are very few observations that one can cross correlate against. Fortunately, this situation is rapidly changing with experiments such as COMAP, CCAT-p and TIME, which are targeting the highly redshifted CO rotational lines and the ionized carbon [CII] fine structure line. Intensity mapping with [CII] can potentially reach $z \sim 9$, enabling direct cross correlations. (CO measurements may be possible too, but there is considerable uncertainty as to whether photodissociations will destroy enough CO molecules at high redshifts to render the signal unobservably small). Since [CII] emission originates from within galaxies, whereas the \tcm line primarily probes the IGM, the two lines provide a complementary view of our Universe; the former probes the \emph{causes} of reionization, whereas the latter probes the \emph{effects}. Moreover, such cross correlations will be crucial in confirming a first detection of the cosmological \tcm signal, since any systematics that are instrument-specific should (on average) vanish in cross correlation.

\subsection{Confirmation and refutation of the EDGES signal} With the unexpected nature of the EDGES result, it is imperative that purported absorption signature be confirmed or refuted. The timing and narrowness of the measured absorption trough implies that star formation rates must evolve much more rapidly at $z> 10$ than expected \citep{MirochaFurlanetto2019}. Moreover, the amplitude of the trough is approximately a factor of 2 larger than is allowable under $\Lambda$CDM unless there exists a previously unknown population of radio loud sources at high redshifts \citep{FengHolder2018,Ewall-Wice2018,Sharma2018,Jana2019,Fialkov2019}. This has generated a large number of theoretical interpretations, including those that involve exotic new physics such as dark matter-baryon scattering \citep{SlatyerWu2018,HiranoBromm2018,Barkana2018}.

\subsection{Fundamental physics with \tcm cosmology} As \tcm experiments move beyond an initial detection and into higher redshift and sensitivity regimes, opportunities will arise to move beyond constraints on the astrophysics of galaxy formation and into constraints on fundamental physics. There are a number of ways to go about this:
\begin{itemize}
\item \textbf{Modelling the astrophysics.} The simplest way to access fundamental physics is to fit the data to parametrized models that include both the astrophysics and the fundamental physics. The weakness of this approach is that it is unclear whether or not our models of Cosmic Dawn and the EoR are good enough to avoid errors. Luckily, a recent breakthrough in the theoretical modelling of reionization has shown that, contrary to previous expectations, ionization fluctuations during the EoR may be efficiently describable using perturbation theory \citep{Hoffmann2018,McQuinnD'Aloisio2018}, enabling \tcm fluctuations to be described as a bias expansion of the matter density field, expressible with a relatively small number of free parameters that can be marginalized out. This provides a statistically disciplined and unbiased way to account for the messy astrophysics of our high-redshift Universe.
\item \textbf{Look for robust signatures.} Another avenue for accessing cosmology is to take advantage of signatures that are expected to be relatively robust in the face of astrophysics. One such possibility is to use redshift space distortions, which are directly sourced by the matter density field and thus bypass the complicated astrophysics governing ionization or spin temperature fluctuations \citep{BarkanaLoeb2005,McQuinn2006}. Early theoretical forecasts of this effect were based on linear theory, but detailed non-linear simulations have recently confirmed the promise of pursuing this type of measurement, at least in the early stages of reionization (when the neutral fraction is less than $\sim 40\%$) \citep{Shapiro2013}. Another intriguing possibility was recently suggested by \citet{ Munoz2019a}, who identified a standard ruler the scale of Velocity-induced Acoustic Oscillations (VAOs), which are sourced by the relative velocities of dark matter and baryons following recombination \citep{TseliakhovichHirata2010}. Instruments like HERA are capable of measuring the VAO standard ruler at $z \sim 15$ to $20$, resulting in percent-level constraints on the Hubble parameter at those redshifts. This opens up kinematical cosmological constratins in an hitherto unexplored redshift regime.
\item \textbf{Combine \tcm measurements with the CMB.} \tcm observations can be used to make predictions for the optical depth to the CMB, $\tau$, which is a crucial ``nuisance" parameter in CMB constraints \citep{Liu2016,FialkovLoeb2016}. This is expected to be a limiting factor in many next-generation CMB measurements taking place over the next decade \citep{CMBS4}. An independent constraint on $\tau$ from \tcm cosmology enables tighter CMB measurements on the sum of the neutrino masses $\sum m_\nu$ (to within $\sim 12\,\textrm{meV}$) or the amplitude of the primordial power spectrum $A_s$ (to better than a percent; \citealt{Liu2016}). Note that CMB constraints on $\sum m_\nu$ are weakened if it is necessary to extend $\Lambda$CDM beyond 6 parameters, increasing the importance of independent constraints \citep{Allison2015}. Of course, our ability to correctly interpret \tcm data in conjunction with CMB data is limited by the correctness of our models for the astrophysics. However, we expect that this situation will improve with ongoing efforts to improve models. Additionally, once the first high-significance measurements of the \tcm line are available, we will be able to \emph{test} rather than \emph{assume} our models of Cosmic Dawn and the EoR.
\end{itemize}

\section{Canadian Involvement in the Current Landscape}
\label{sec:CurrentLeader}
In the last decade, the Canadian community has played key leadership roles in \tcm cosmology at $z > 6$.
\begin{itemize}
    \item \textbf{Earliest upper limits.} The earliest upper limits on spatial fluctuations at $z > 6$ were led by former Canadian Institute for Theoretical Astrophysics (CITA) grad student Dr. Gregory Paciga and CITA faculty member Prof. Ue-Li Pen \citep{Paciga2013}.
    \item \textbf{Leadership in international collaborations.} The Canadian community plays a leadership role in multiple instruments that are expected to detect and characterize \tcm spatial fluctuations in the next few years. Via the sponsorship of Prof. Bryan Gaensler (University of Toronto and Dunlap Institute), Canadian scientists are able to be a part of the MWA collaboration. McGill University is a full institutional partner of the HERA collaboration via Prof. Adrian Liu, who as the Power Spectrum Lead Scientist is responsible for the delivery of the project's main goal of measuring the power spectrum of spatial \tcm fluctuations. Finally, a large number of Canadian scientists remain in leadership positions within the Square Kilometre Array (SKA) Organization, enabling ``lessons learned" from current instruments to be (relatively) easily incorporated into the design and construction of the SKA.
    \item \textbf{Common technology with \tcm cosmology at $z<6$.} With the Canadian Hydrogen Intensity Mapping Experiment (CHIME; \citealt{Bandura2014}) and the Hydrogen Intensity and Real-time Analysis eXperiment (HIRAX; \citealt{Newburgh2016}), Canada is the undisputed leader of \tcm cosmology at $z<6$. Many of the technologies developed for this effort (e.g., digital correlators for large radio interferometers) as well as many of the analysis tools (e.g., high-precision foreground mitigation algorithms) are essentially identical to the ones that are needed at $z>6$.
    \item \textbf{A diverse portfolio of global 21-cm experiments.}  Canada has extensive involvement in a variety of experiments that are targeting the globally averaged 21-cm signal.  In addition to collaborative ties with EDGES, Canadian reseachers are also playing key roles in Shaped Antenna measurement of the background RAdio Spectrum (SARAS; \citealt{Saurabh2018}), Mapper of the IGM Spin Temperature (MIST), and Probing Radio Intensity at high-Z from Marion (PRI$^{\rm Z}$M; \citealt{Philip2018}).  In particular, the PRI$^{\rm Z}$M leadership is based at McGill through Profs. Cynthia Chiang and Jonathan Sievers.  The above projects comprise over half of the global 21-cm experimental efforts worldwide.
    \item \textbf{Complimentary measures of EoR with the [CII] line.} Canada has involvement in one of the major current experiments to probe [CII], with the PI of TIME Abigail Crites joining the University of Toronto and the Dunlap Institute.   Complimentary experimental probes of EoR will be key for cross-correlations which will allow a more robust detection of cosmology during EoR and provide additional scientific information about the bubble size during EoR.
\end{itemize}

\section{Opportunities and Recommendations for Canada in the coming decade}

\begin{itemize}
    \item  \textbf{Continued strong investment in $\mathbf{z < 6}$ experiments.}  In the large $n_\textrm{ant}$ limit, both high- and low-$z$ experiments use similar calibration and analysis techniques.  The $m$-mode framework for optimal mapmaking developed by CHIME \citep{Shaw2014} was used to make the first all-sky maps from the LWA \citep{Eastwood2018}.  Redundant calibration \citep{Liu2010} and derivatives thereof have become the \textit{de facto} tools to calibrate both low- and high-$z$ arrays, with PAPER, HERA, CHIME, and HIRAX all pursuing similar calibration strategies.  As a result, new tools developed by one experiment are immediately useful to the others.  Understanding and removing foregrounds is the single largest challenge for these arrays, and so there is synergy in the  increased spectral coverage from combining low- and high-$z$ data.  Similarly, effects of the ionosphere are a challenge for high-$z$ arrays as well as for  sub-arcsecond localization of Fast Radio Bursts (FRBs) that will be carried out by high-frequency arrays with outriggers.  While the effects are much larger at low frequencies due to the $\lambda^2$ scaling of plasma dispersion, high-frequency arrays have access to global navigational satellite system (GNSS) frequency bands, which provide high SNR, and hence high time resolution, measurements of ionospheric delays through a limited set of pierce points.  The combination of ionospheric monitors will have more information then either redshift range by itself, which can benefit both redshift ranges.  Of course, coordination would be required to take maximal advantage.
    
    \item \textbf{Continue investments in current and next-generation $z > 6$ experiments.} Historically, Canadian investment in \tcm spatial mapping efforts have been limited to the $z< 6$ regime. However, as discussed in Section \ref{sec:CurrentLeader}, Canadian astronomers have nonetheless led international collaborations from leadership positions oriented towards analysis and theory. This is possible because spatial mapping experiments are in many ways software telescopes, where precision analysis techniques are just as important as the hardware in making the experiments a success. Continued Canadian leadership is therefore possible with key analysis and theory investments over the next decade in international efforts such as the MWA, HERA, and SKA. In support of this and in recognition of the importance of data techniques, we also recommend continued investment in data-oriented efforts such as the Canadian Initiative for Radio Astronomy Data Analysis (CIRADA).
    \item \textbf{Strong investment in global signal experiments.}  Canada is already playing a leading role in measurements of the globally averaged 21-cm signal through through multiple experimental efforts. The reported EDGES detection represents our first view into Cosmic Dawn, an epoch that is ripe for new exploration.  Future instruments aiming to make competitive measurements at these frequencies must operate from 1) remote locations where RFI is minimized, and 2) locations with quiet ionospheric conditions---polar latitudes, especially at night during solar minima, are excellent candidates.  The high Arctic meets both of these criteria and presents a unique Canadian geographic advantage that may allow us to open a brand new window on the radio sky.
    \item \textbf{Lay the groundwork for future exploration of the Dark Ages.} The cosmic Dark Ages are unexplored to date and represent one of the final observational frontiers in cosmology.  This epoch contains a potential wealth of cosmological information~\citep{LoebZaldarriaga2004}, but observations at frequencies of $\lesssim 30$~MHz are exceptionally difficult because of bright foreground emission, interference from the ionosphere, and contamination from terrestrial RFI.  At the very lowest frequencies, the state of the art among ground-based measurements dates from the 1950s, when Reber and Ellis caught brief glimpses of the $\sim 2$~MHz sky at low resolution \citep{Reber1956}. Despite the present-day RFI environment, low-frequency observations may still be accessible from carefully selected locations and with new technology developments.  Preliminary observations from Marion Island show that clear interferometric fringes from astronomical sources are visible down to $\sim$10~MHz.  Canada has both the leadership and expertise needed to push these measurements to the next level and begin taking the first steps toward future exploration the Dark Ages.
    \item \textbf{Invest in complementary line intensity mapping experiments that deepen our understanding of EoR and complement the 21cm probe.} To reap the benefits from the cross-correlations especially between [CII] and 21cm, a large [CII] effort will be necessary to match the current scales of 21 cm efforts in the coming decade.  To continue to be a leader in this emerging field, Canada should invest in a large [CII] experimental effort especially in terms of receiver cameras and investment or partnership in a large sub-mm telescope at a sub-mm quality site (e.g. the Atacama or Antarctica). 
    \item \textbf{Continue investment in theoretical efforts.} The Canadian theoretical community has played a key role in providing models, making predictions, running simulations, and proposing observational signatures for \tcm cosmology. Examples of this include (but are not limited to) some of the largest-scale simulations of reionization in the world \citep{alvarez/abel:2012} (pioneered by former CITA postdoc Marcelo Alvarez), interpretations of surprising recent Lyman-$\alpha$ forest data and their implications for reionization (led by current CITA postdoc Laura Keating), proposals for how \tcm surveys can be combined with intensity mapping of other lines to yield greater science returns (by current CITA postdoc Patrick Breysse), and constraints on high-redshift luminosity functions implied by the EDGES result (by current McGill CITA National Fellow Jordan Mirocha). Robust theory investments are crucial for a successful high-redshift \tcm program, and we strongly suggest continued investment in theory programs such as CITA.
	\item \textbf{Maintain leadership opportunities and knowledge transfer to international partners such as the SKA.} The lessons learned from Canadian involvement in current-generation instruments will be invaluable for experiments coming to the forefront in the later part of the next decade, such as the SKA. Current Canadian efforts will strongly reduce risk in projects like the SKA, and thus it is crucial to maintain links with international efforts. For the SKA in particular, Canada is particularly well-suited to provide technical leadership, for while both the US and Canada are learning tremendous amounts from tightly intertwined North American efforts, the US (unlike Canada) has no formal links to the SKA. We are therefore poised to fill a critical hole in international knowledge transfer.
\end{itemize}




\begin{lrptextbox}[How does the proposed initiative result in fundamental or transformational advances in our understanding of the Universe?]
High-redshift \tcm cosmology opens up an entirely new redshift regime to direct observations. This provides access to a broad range of previously unexplored astrophysical and cosmological phenomena, particularly those connected to the formation of the first stars and galaxies. Given that key properties of these first luminous objects are currently not known to within an order of magnitude, this would represent a significant advance in our understanding of our Universe. Moreover, as evidenced by the tentative EDGES result, there is significant room for surprises and unexpected physics beyond the standard paradigms. Indeed, one could argue that high-redshift \tcm cosmology amounts to a set of high-precision measurements in low-frequency radio, which is a band that has historically been poorly surveyed. The potential for discovery is therefore high, given that historically, the opening up of new wavelengths to observations has always resulted in unexpected phenomena.
\end{lrptextbox}

\begin{lrptextbox}[What are the main scientific risks and how will they be mitigated?]
The biggest scientific risk is the technical difficulty of \tcm cosmology. In particular, the problems of foreground contamination, systematics control, and sensitivity have meant that a \emph{definitive} first detection of the high-$z$ \tcm signal has been elusive.\footnote{Until the EDGES signal is independently verified, we will be conservative and say that there is still work to be done before a robust detection can be declared.} However, there are ways in which this risk can be mitigated. First, it is important to note that \tcm experiments have now crossed into a new regime where there is an abundance of science-grade data available for analysis. Exercises such as the removal of systematics are no longer hypothetical ones, and the data are already providing important lessons for future analyses and future instruments. Second, cross correlations with other lines should provide a cleaner path towards a positive detection, given that independent systematics should vanish (on average) in a cross-correlation analysis. This underscores the importance of investing in instruments that perform intensity mapping with lines complementary to the \tcm line.
\end{lrptextbox}

\begin{lrptextbox}[Is there the expectation of and capacity for Canadian scientific, technical or strategic leadership?] 

Canadian astronomers are well-positioned to maintain and expand their leadership roles in \tcm cosmology. Gaensler and Liu are part of the MWA and HERA executive boards, respectively. Chiang and Sievers are the PIs of PRI$^\textrm{Z}$M, and Crites is the PI of the TIME experiment. The Canadian community also provides scientific leadership despite the fact that many high-redshift \tcm experiments are located out of Canada. This is possible because in many ways the relevant telescopes are \emph{software} telescopes, where high-precision analyses are just as important as high-precision hardware. Many members of the Canadian community are deeply embedded in the current hardware, analysis, and theory efforts in the US. These efforts are not only scientifically interesting in their own right, but may also inform the SKA project. This places Canada in a unique position for strategic leadership, given that the US has no formal involvement with the SKA.

\end{lrptextbox}

\begin{lrptextbox}[Is there support from, involvement from, and coordination within the relevant Canadian community and more broadly?] 
The strong Canadian presence in the \emph{low} redshift \tcm landscape (via projects like CHIME, HIRAX, and CHORD), provides significant support, involvement, and coordination within the Canadian community. For example, most of of the scientists engaged in \emph{high} redshift \tcm science are also part of a low-redshift effort, enabling relatively easy coordination and support (including but not limited to technical support with shared techniques and instrumentation). In addition, many experiments targeting complementary lines have cross correlation as an explicit goal in their design, highlighting the coordination of the relevant communities. Finally, Canada has the infrastructure to support data analysis-intensive ventures like \tcm cosmology, with data centres such as CIRADA.

\end{lrptextbox}

\begin{lrptextbox}[Will this program position Canadian astronomy for future opportunities and returns in 2020-2030 or beyond 2030?] 
High-redshift \tcm cosmology will continue to provide opportunities throughout the next decade and beyond. Beyond the opportunities outlined in this proposal, future instruments will push to yet higher redshifts, finer scales, and greater sensitivities. This will enable even higher precision explorations of the nature of the first stars and galaxies, and potentially even lead to observations of the treasure trove of cosmological modes at truly high redshifts, i.e., the Dark Ages. The technical knowledge accrued from obtaining a first detection of the highly redshifted \tcm signal will be invaluable for this effort, and will mitigate risk in future experiments. A prime example of this would be the second phase of the SKA, which will become increasingly relevant as one approaches 2030.
\end{lrptextbox}

\begin{lrptextbox}[In what ways is the cost-benefit ratio, including existing investments and future operating costs, favourable?] 
In general, \tcm cosmology is an extremely cost-effective venture, requiring relatively inexpensive hardware for a large-scale cosmology experiment that is designed to map spatial fluctuations. Moreover, the extreme importance of software/analysis in \tcm cosmology means that relatively inexpensive investments in Canadian analysis teams can produce a magnified influence on international experiments. In addition, global signal experiments provide an avenue for small research groups (at the individual university/PI level) to have a large impact on the field.
\end{lrptextbox}

\begin{lrptextbox}[What are the main programmatic risks
and how will they be mitigated?] 
While many \tcm efforts have been funded for the next few years, the situation is less certain for experiments targeting complementary lines such as [CII] or CO. Many such experiments are funded for just their pilot phases, rather than the full instruments (and surveys) needed for high-significance cross correlation measurements. An additional risk is the fact that large-scale spatial mapping \tcm experiments are primarily funded by non-Canadian sources, and thus the availability of appropriate observational facilities beyond the next few years will depend on the results of processes like the US Decadal Survey.
\end{lrptextbox}

\begin{lrptextbox}[Does the proposed initiative offer specific tangible benefits to Canadians, including but not limited to interdisciplinary research, industry opportunities, HQP training,
EDI,
outreach or education?] 
Hardware experience is becoming increasingly difficult for HQP to come by, given the trend towards larger facilities and large collaborations. The fact that \tcm cosmology is a relatively young field means that many experimental efforts (e.g., global \tcm signal experiments and non-\tcm intensity mapping experiments) are still relatively small hardware projects that can be done in-house by a faculty member's research group within an HQP's time with a group. On the software/analysis side, the large datasets that are routinely generated by \tcm experiments provide an attractive venue for interdisciplinary research with computer scientists, particularly those that come from the machine learning community. Finally, the fact that many experiments are located at remote sites such as South Africa provides an opportunity for outreach to communities that Canadians would otherwise not have contact with.
\end{lrptextbox}

\bibliography{example} 

\end{document}